\documentclass[epj]{svjour}
\usepackage{graphics}
\usepackage{graphicx}
\usepackage{dcolumn}
\usepackage{bm}
\usepackage{CJK}
\usepackage[usenames]{color}
\usepackage{amsmath}
\begin{document}
\begin{CJK*}{GB}{gbsn}
\title{Nucleon-number scalings of anisotropic flows and nuclear modification factor for light nuclei in the squeeze-out region}
\author{T. T. Wang \inst{1,2} \and Y. G. Ma \inst{1,3} \thanks {Email: mayugang@fudan.edu.cn}}
%
\offprints{}          
\institute{Shanghai Institute of Applied Physics, Chinese Academy of Sciences, Shanghai 201800, China \and University of Chinese Academy of Sciences, Beijing 100049, China \and Key Laboratory of Nuclear Physics and Ion-Beam Application (MOE), Institute of Modern Physics, Fudan University, Shanghai 200433, China}
\date{Received: date / Revised version: date}
%
\abstract{The number of nucleon (NN) scaling of  the directed flow $v_1$ and elliptic flow ($v_{2}$) as well as the nuclear modification factor ($R_{cp}$) are tested for light nuclei which are produced in 0.4$A$ GeV $^{197}Au + ^{197}Au$ collisions at different impact parameters  with two different in-medium nucleon-nucleon cross sections in the framework of an isospin-dependent quantum molecular dynamics (IQMD) model. 
In that energy domain, the emission of light nuclei can be well described by the squeeze-out phenonomenon. The results show a nice NN scaling behavior for flow parameters $v_{1}$, $v_{2}$ and ${R_{cp}}$. 
These results demonstrate that 
the nucleon coalescence mechanism is responsible for nucleon-number scaling of above physical observables 
in squeeze-out region in heavy-ion collisions at intermediate energy. }


\maketitle

\section{Introduction}\label{sec.I}

In intermediate-energy heavy-ion collisions (HIC), the collective flow is one of the important observables which can reflect the early stage of heavy ion reaction, evolution dynamics,  in-medium nucleon-nucleon cross section~\cite{b1,b2,b3,b4,b5,b7,b10}, as well as the information on the nuclear equation of state (EOS)~\cite{b11,b12,b13,b14,b15,b16,b17,b18,b19}.
Many studies of anisotropic flows, namely the directed flow ($v_{1}$) and the elliptic flow ($v_{2}$) revealed the properties and origin of the collective motion through exploring the dependences for anisotropic flows on beam energy, fragment mass, impact parameter and so on in HIC~\cite{b11,b12,b13,b14,b15,b16,b17,b18,b19}.

The number-of-nucleon (NN) scaling of  anisotropic flows was first proposed by Yan and Ma {\it et al.} in low-intermediate energy heavy ion collisions through the framework of a quantum molecular dynamics model~\cite{b20,b21}.
This scaling states that when the anisotropic flows were scaled by the nucleon-number of light nuclei, these scaled flows will follow the same curve as a function of rapidity or transverse momentum per nucleon regardless of the mass of light nuclei
~\cite{b21}. 
This phenomenon  can be naturally  interpreted by the nucleonic coalescence mechanism for the formation of light nuclei at kinetic freeze-out stage of the reaction system. However,  the earlier focus of the coalescence mechanism is only on the spectra of kinetic energy or momentum of light nuclei in heavy ion collisions~\cite{b29} but not on flow behavior.  
After the first prediction on number of nucleon scaling of  flows in  heavy-ion collisions at lower energies~\cite{b20,b21}, 
the same concept of number-of-nucleon scaling of elliptic flow was  followed by a dynamical model as well as a multiphase transport (AMPT) model calculations at ultra-relativistic energy ~\cite{b24,b25,b30}. 
 Later on the number-of-nucleon scaling of  anisotropic flows for light nuclei have been experimentally confirmed in Au + Au collisions at much higher energies, such as 200, 62.4, 39, 27, 19.6, 11.5, and 7.7 GeV at  the BNL Relativistic Heavy Ion Collider (RHIC)~\cite{b22}, which illustrates that the light nuclei at RHIC energies are also formed by nucleonic  coalescence mechanism at the freeze-out stage but not directly  by the quark coalescence mechanism.
 At both lower and relativistic energies, the elliptic flow demonstrates the positive values: the former stems from the collective rotational motion due to the dominated mean field, and the later originates from the strong pressure of the overlapping participant region. In both cases, the NN scaling has been examined. However, in the intermediate-energy regime where  the negative elliptic flow might emerge  due to the squeeze-out behavior of particle emission, NN scaling of flows has not yet checked so far.

On the other hand,  the number-of-constituent-quark scaling of nuclear modification factor  (NMF) for pions and protons as well as the number-of-nucleon  scaling of NMF for light nuclei have been first proposed by one of the authors~\cite{b31}, and they were experimentally supported for Au + Au and Pb + Pb collisions at NA49, RHIC and LHC energies.  However, the similar NN scaling behavior of NMF is not yet examined in intermediate energy heavy ion collisions.  In this  energy domain, the isospin-dependent quantum molecular dynamics (IQMD) model ~\cite{b27,b28,b32} is very successful for studying reaction dynamics, such as for spectra analysis, flow analysis as well as  the nuclear modification factor analysis~\cite{b33}. 
Therefore it will be of great interest to test if the ${R_{cp}}$ of light nuclei  can be scaled  by the number of nucleon scaling.

Based upon the above arguments,  in this article, we use an isospin-dependent molecular dynamics (IQMD) model to simulate 0.4$A$ GeV Au + Au collisions 
at different impact parameters with two sets of in-medium nucleon-nucleon cross section. The anisotropic flows and $R_{cp}$ of different light nuclei especially in terms of the coalescence mechanism are systematically investigated, and  the squeeze-out behavior was clearly observed by the negative elliptic flow. In this work the ${R_{cp}}$ is obtained by dividing the spectra in the centrality of 0-10$\%$, 10-20$\%$, and 20-40$\%$ to the one in the centrality of 60-80$\%$. Dependences of the mass number of the anisotropic flows and the nuclear modification factor are surveyed, and the nucleon-number scaling of light nuclei on the anisotropic flows and $R_{cp}$ are also found.

\section{Theoretical descriptions}\label{sec.II}
\subsection{Definition of the directed and elliptic flows}\label{sec.IIA}

Anisotropic flow is defined as the different $n$-th harmonic coefficient $v_{n}$ of the Fourier expansion of the particle invariant azimuthal distribution
\begin{equation}
\label{eq1} 
\frac{d N}{d \phi} \propto 1+ 2 \sum_{n=1}^{\infty} v_{n} \cos (n \phi),
\end{equation}
where the reaction plane was defined by a plane of the z-axis along the beam axis and x-axis along the impact parameter axis in the coordinate system and $\phi$ is the azimuthal angle between the transverse momentum of the particle and the reaction plane.
For a given rapidity the anisotropic flows 
are evaluated according to
\begin{equation}
\label{eq2} 
v_{n}\left(p_{t}\right) = \langle\cos (n \phi)\rangle, 
\end{equation}
where $\langle\cos (n \phi)\rangle$ denotes the average over the azimuthal distribution of particles with transverse momentum $p_{t}$. 
Specifically, the directed flow 
\begin{equation}
\label{eq3} 
v_{1} = \langle\cos \phi\rangle=\left\langle\frac{p_{x}}{p_{t}}\right\rangle,
\end{equation}
and elliptic flow 
\begin{equation}
\label{eq4} 
v_{2}=\langle\cos(2\phi)\rangle=\left\langle\frac{p_{x}^{2}-p_{y}^{2}}{p_{t}^{2}}\right\rangle, 
\end{equation}
where  $p_{x}$ and $p_{y}$ are components of transverse momentum, and  $p_t = \sqrt{p_{x}^{2} + p_{y}^{2}}$ is transverse momentum.

\subsection{Definition of nucleon-number scaling of nuclear modification factor}\label{sec.IIB}

The nuclear modification factor ${R_{cp}}$ is a probe to study the nuclear medium effect in  nucleus-nucleus collisions, it is defined as a ratio of the particle yield in central collisions to that in peripheral collisions ~\cite{b31,b33,b34}, i.e. 
\begin{equation} 
\label{eq5}
R_{cp} = \frac{\left[d^{2} N / 2 \pi p_{T} d p_{T} d y /\left\langle N_{bin}\right\rangle\right]^{central}}{\left[d^{2} N / 2 \pi p_{T} d p_{T} dy /\left\langle N_{bin}\right\rangle\right]^{peripheral}}, 
 \end{equation}
where $d^{2} N / 2 \pi p_{T} d p_{T} d y$ is the particle invariant differential yield at the transverse momentum spectra $p_{T}$, $\left\langle N_{bin}\right\rangle$ means the average nucleon-nucleon binary collision number per event.
Now, we focus on  the coalescence mechanism for the formation of light nuclei similar to that for the production ~\cite{b35} or elliptic flow~\cite{b3,b36} of hadrons in relativistic heavy-ion collisions. The formation of light nuclei can be described by the coalescence mechanism as follows:
\begin{equation} 
\label{eq6}
\begin{array}{l}{E_{A} \frac{d^{3} N_{A}}{d^{3} P_{A}} = B_{A}\left(E_{p} \frac{d^{3} N_{p}}{d^{3} P_{p}}\right)^{Z}\left(E_{n} \frac{d^{3} N_{n}}{d^{3} P_{n}}\right)^{A-Z}} \\    
      = {B_{A}\left(E_{p} \frac{d^{3} N_{p}}{d^{3} P_{p}}\right)^{A}}, 
\end{array}
\end{equation}
where $P_{p} = P_{n} = P_{A}/A$ and $A$ is the number of nucleons in a light nucleus. The coefficient $B_{A}$ is the probability for the coalescence production of a nucleus with mass number $A$, which depends on momentum as well as the fireball volume in coordinate space~\cite{b37}. The coefficient $B_{A}$ can be extracted from data or calculated by the coalescence mechanism~\cite{b37}. The number of nucleon scaling for the ${R_{cp}}$ of light nuclei can be deduced from Eq.~\ref{eq5} and Eq.~\ref{eq6}  ~\cite{b31} :
\begin{equation}
\label{eq7} 
R_{cp}^{*} = \left(\frac{B_{A,c}}{B_{A,p}}\right)^{-1 / A}\left(R_{c p}\left(A\cdot p_{T}\right)\right)^{1 / A}\left(\frac{\left\langle N_{b i n}\right\rangle^{c}}{\left\langle N_{b i n}\right\rangle^{p}}\right)^{1 / A-1}.
\end{equation}
Here, the index $c$ and $p$ stands for the central collisions and peripheral collisions, respectively. $R_{cp}^{*}$ is the scaled nuclear modification factor according to the scaled proton $R_{cp}$. Then we can try to scale $R_{cp}$  of  light nuclei by the following formula ~\cite{b31}:
\begin{equation} 
\label{eq8}
\widetilde{R}_{cp}^{*}\left(p_{T}\right) = \left(R_{c p}\left(A \cdot p_{T}\right)\right)^{1 / A}\left(\frac{\left\langle N_{b i n}\right)^{c}}{\left\langle N_{b i n}\right\rangle^{p}}\right)^{1 / A-1},
\end{equation}
and 
\begin{equation}
\label{eq9} 
\widetilde{R}_{cp}^{*}\left(p_{T}\right)=R_{c p}^{*}\left(p_{T}\right) /\left(\frac{B_{A,c}}{B_{A,p}}\right)^{-1 / A}.
\end{equation}

\subsection{The isospin-dependent quantum molecular dynamics model}\label{sec.IIC}

In the following discussion, we introduce the isospin-dependent quantum molecular dynamics (IQMD) model briefly. The quantum molecular dynamics model is a n-body transport theory, it describes heavy-ion reactions from intermediate to relativistic energies. The main parts of QMD transport model include the following: the initialization of the projectile and the target, nucleon propagation in the effective potential, the collisions between the nucleons in a nuclear medium, the Pauli blocking effect, and the numerical test.
The isospin-dependent quantum molecular dynamics model is based on the QMD transport model with the isospin factors taken into account~\cite{b27,b38}. As we know, the main components of the dynamics in heavy-ion collisions (HICs) at intermediate energies include the mean field, two-body collisions, and Pauli blocking. Therefore, it is important for these three components to include isospin degrees of freedom in the IQMD transport model. 

In particular,  the density-dependent Skyrme potential $U_{Sky}$   reads when the momentum dependent potential is included
\begin{multline}
\label{eq10}
 U_{\mathrm{Sky}}= \alpha\left(\frac{\rho}{\rho_{0}}\right)+\beta\left(\frac{\rho}{\rho_{0}}\right)^{\gamma} \\ 
 +\frac{\rho}{\rho_{0}} \int d \vec{p}^{\prime} g\left(\vec{p}^{\prime}\right) \delta \ln ^{2}\left[\epsilon\left(\vec{p}-\vec{p}^{\prime}\right)^{2}+1\right],
\end{multline}
where $\rho$ and $\rho _{_0}$ are total nucleon density and its
normal value, respectively. Here, $g(\vec{p}, t)= \frac{1}{(\pi \hbar)^{3 / 2}} \sum_{i} e^{-\left[\vec{p}-\vec{p}_{i}(t)\right]^{2} \frac{2 L}{(\hbar)^{2}}}$ is the momentum distribution function. The parameters $\alpha$, $\beta$,
$\gamma$, $\delta$ and $\varepsilon$ are related to the nuclear
equation of state~\cite{b39,b40} and listed in Table I, 
 where $K$ = 200 or 380 MeV means  the soft- or
the stiff-momentum dependent potential, respectively.

\begin{table}[!htbp]
\caption
{The parameters of the interaction potentials.}
\begin{tabular}{|c|c|c|c|c|c|c|c|}
\hline
$\alpha$ & $\beta$&$\gamma$&$\delta$& $\varepsilon$ & K\\
\hline (MeV)&(MeV)&&(MeV)&(MeV)&(MeV)\\\hline
-390.1&320.3&1.14&1.57&21.54&200\\\hline
-129.2 &59.4&2.09 &1.57&21.54&380\\\hline
\end{tabular}\\
\end{table}

As well known, the dynamical coalescence model has been  used  for describing the production of light clusters in heavy-ion collisions~\cite{b39-1} at both intermediate and high energies~\cite{b39-2,b39-3,b39-4,b39-5,b39-6,b39-7,b39-8}. As one of several approaches for the coalescence calculations, the Wigner function approach is discussed by using an extension to the transport theoretical approach RQMD in the literature~\cite{b39-6}. For microscopic coalescence models using the deuteron and triton Wigner functions,  it can also be applied in the quantum molecular dynamical model, especially for treating the absolute yield of particles' spectra. Considering that we just discuss the collective flows and yield ratios in this work, here we will not discuss dynamical coalescence model in this article. Instead, the particles are identified using a modified isospin-independent coalescence description, i.e., two nucleons are assumed to belong to the same cluster if their centers are closer than a distance of 3.5 fm and their relative momentum smaller than 0.3 GeV/c. If the nucleon is not bounded by any clusters, it is treated by an emitted (free) nucleon. There are already many literatures using the separation parameters of 3.5 fm in coordinate space and 0.3 GeV in momentum space for nucleons to form a light nucleus, which has given a good agreement with the experimental data~\cite{b40-1,b40-2,b45,GuoCC}. In the IQMD model, the in-medium nucleon-nucleon cross section (NNCS) 
 is represented by the formula:
\begin{equation}
\label{eq11}
\sigma_{NN}^{med} = \left(1-\eta\frac{\rho}{\rho_{0}}\right)\sigma_{NN}^{free},
\end{equation}
where $\eta$ is the in-medium NNCS reduction factor and $\sigma_{NN}^{free}$ is the available experimental NNCS from Particle Data Group. The above in-medium NNCS reduction factor $\eta$ varied from 0 to 1 and the relationship between $\eta$ and the in-medium effect is opposite in this expression. In particular, the factor $\eta \approx 0.2$ has been found better to reproduce the flow data.

\section{Results and discussion}\label{sec.III}

\begin{figure*}[!htb]
\includegraphics[width=\hsize]{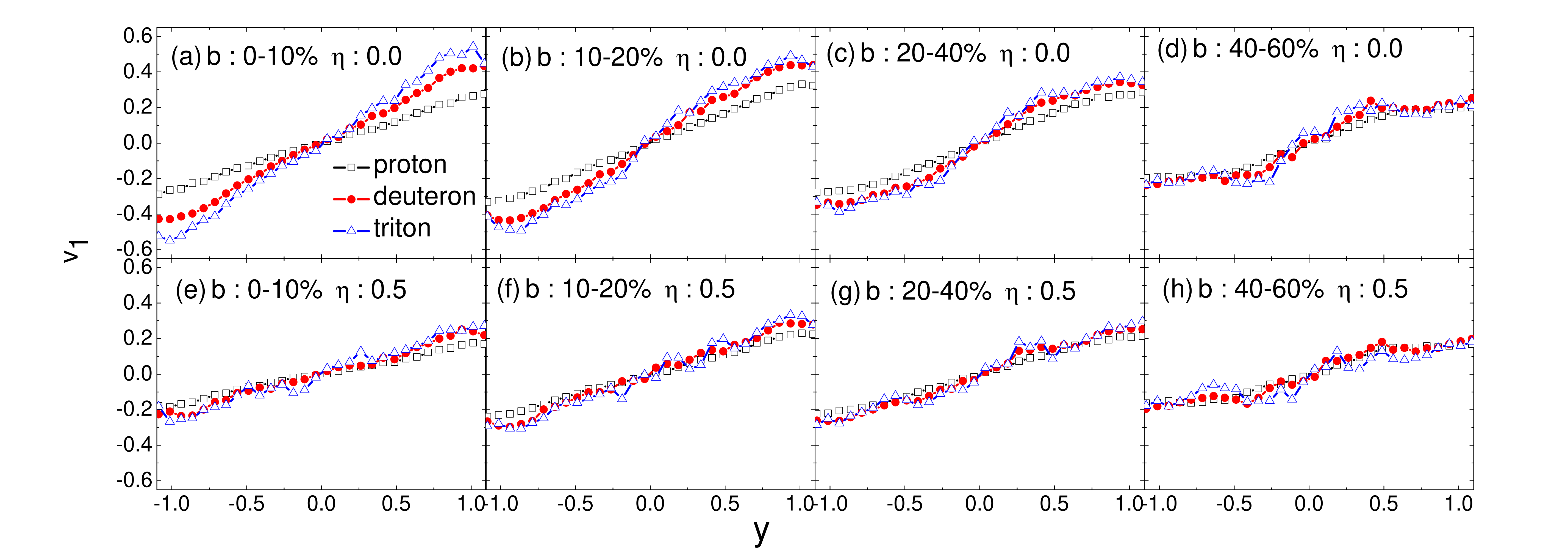}
\caption{(Color online) Rapidity dependence of ${v_{1}}$ for light nuclei for  0.4$A$ GeV Au + Au collisions at different centralities. Squares represent for protons, circles for deuterons, up-triangles for tritons. From the left panel to right panel, it corresponds to different centralities of 0 - 10$\%$,  10 - 20$\%$,  20 - 40$\%$ and  40 - 60$\%$. The upper row represents the calculations with a factor $\eta$ of the in-medium nucleon-nucleon cross section 0, and the lower row with $\eta$ = 0.5.  }
\label{fig1}
\end{figure*}

\begin{figure*}[!htb]
\includegraphics[width=\hsize]{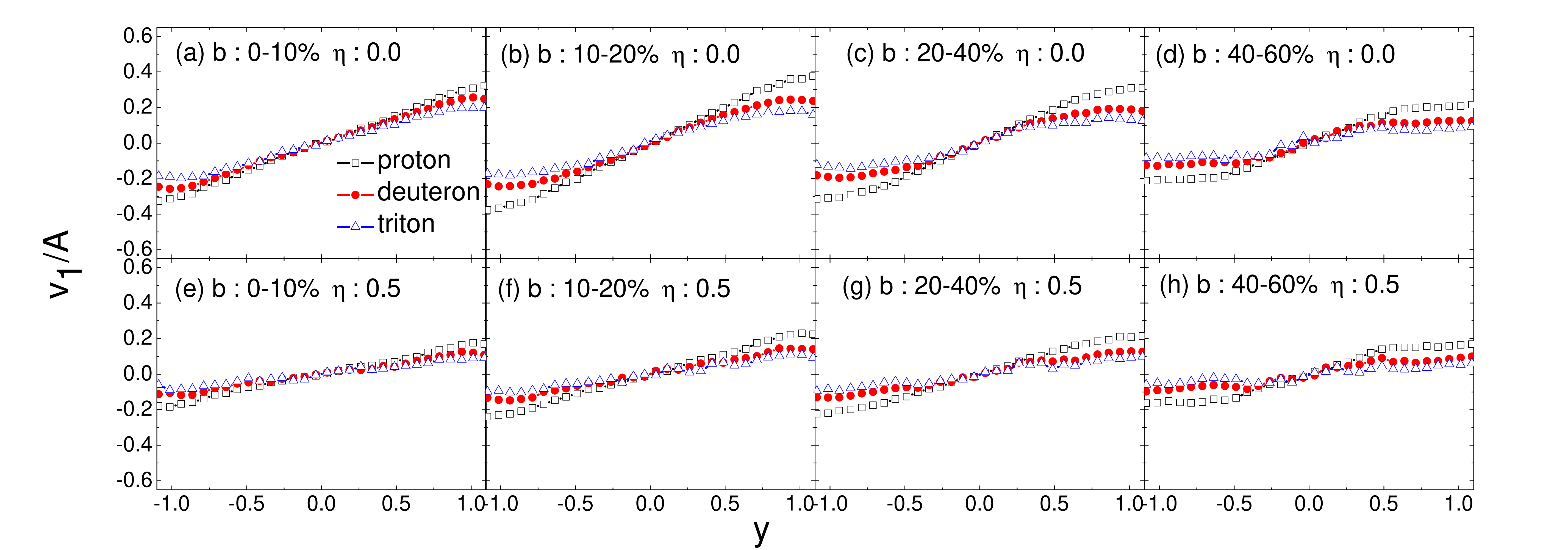}
\caption{(Color online) Same as Fig.~1 but for the  ${v_{1}/A}$ as a function of rapidity.
}
\label{fig2}
\end{figure*}

Reactions for the collision system of ${^{197}Au} + {^{197}Au}$ at 0.4$A$ GeV with different impact parameters have been simulated by IQMD model under  the soft EOS with momentum-dependent interaction (SM+MDI). In this work, the physics results are extracted at 200 fm/c when the reaction system reaches the freeze-out stage. 
Fig.~\ref{fig1} and Fig.~\ref{fig2} shows the anisotropic flows $v_{1}$ as a function of rapidity and the flows per nucleon $v_{1}/A$ versus rapidity for testing the effect of the number of nucleon scaling, respectively, at different centralities with two sets of in-medium nucleon-nucleon cross section. In Fig.~\ref{fig1}, we show the trend of the directed flow $v_{1}$. We found that the directed flow (slope in mid-rapidity) is positive,  
indicating the repulsive nucleon-nucleon interaction are dominant,  which is opposite to the negative slope $v_{1}$ at low energy~\cite{b21} where the attractive mean field plays a key role. It also shows the absolute values of directed flow are larger for heavier clusters at a given rapidity. On the other hand, with the increase of NNCS reduction factor $\eta$ from the upper panels to lower panels in Fig. 1, the directed flow values seem to drop, indicating the origin of flow most comes from the nucleon-nucleon collisions. After we perform the NN scaling of $v_1$, we display Fig.~\ref{fig2} where the curves for different clusters almost stay together around mid-rapidity by the nucleon-number scaled $v_{1}/A$, which illustrates  that the directed flow of the light clusters satisfies the number-of-nucleon scaling. 

\begin{figure*}[!htb]
\includegraphics[width=\hsize]{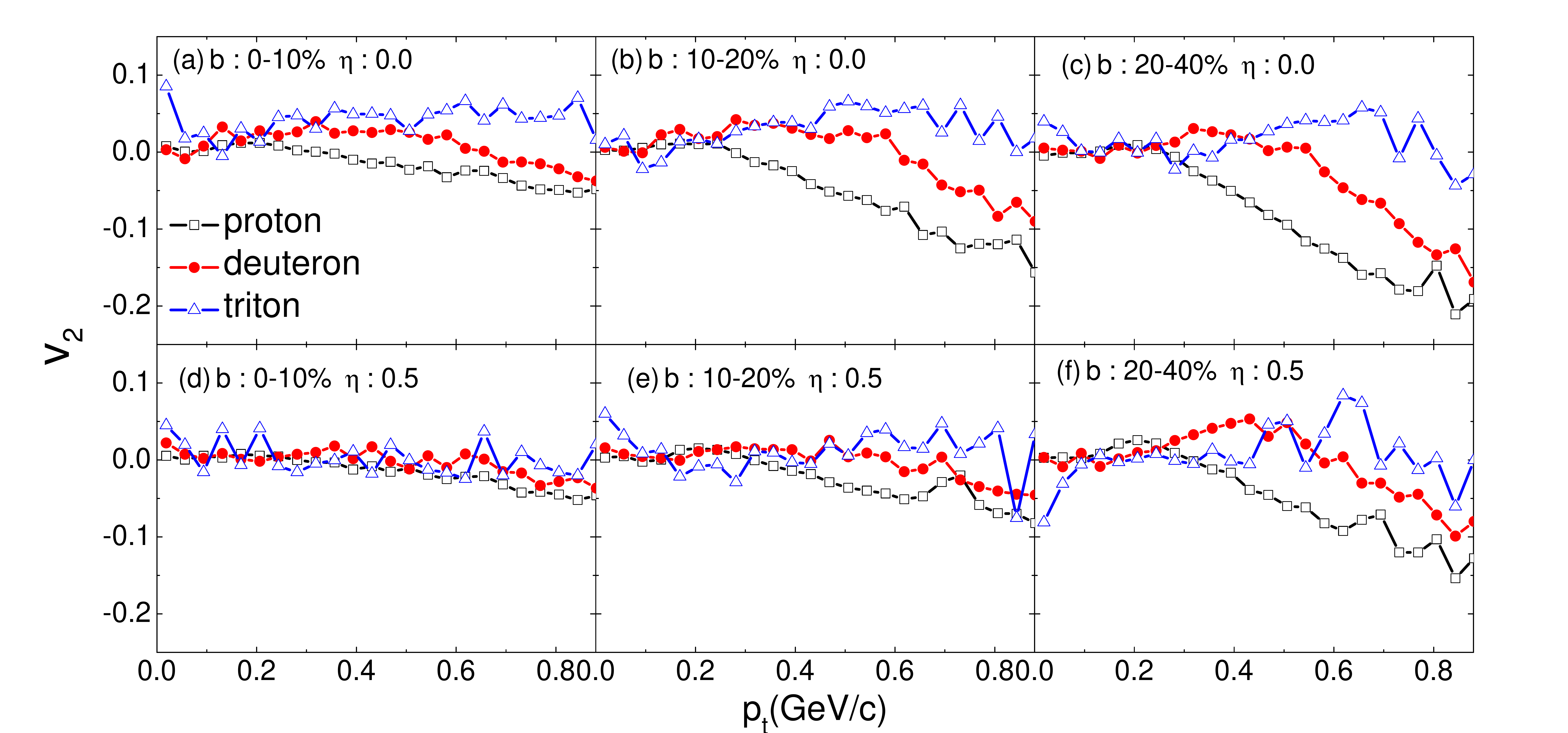}
\caption{(Color online)
The transverse momentum (${p_{t}}$) dependence of ${v_{2}}$ for light nuclei in 0.4 $A$ GeV Au + Au collision at different centralities (from left to right: 0 - 10$\%$,  10 - 20$\%$,  20 - 40$\%$). Squares represent for protons, circles for deuterons, triangles for tritons. The upper row represents the calculations with a factor $\eta$ of the in-medium nucleon-nucleon cross section 0, and the lower row with $\eta$ = 0.5. }
\label{fig3}
\end{figure*}

\begin{figure*}[!htb]
\includegraphics[width=\hsize]{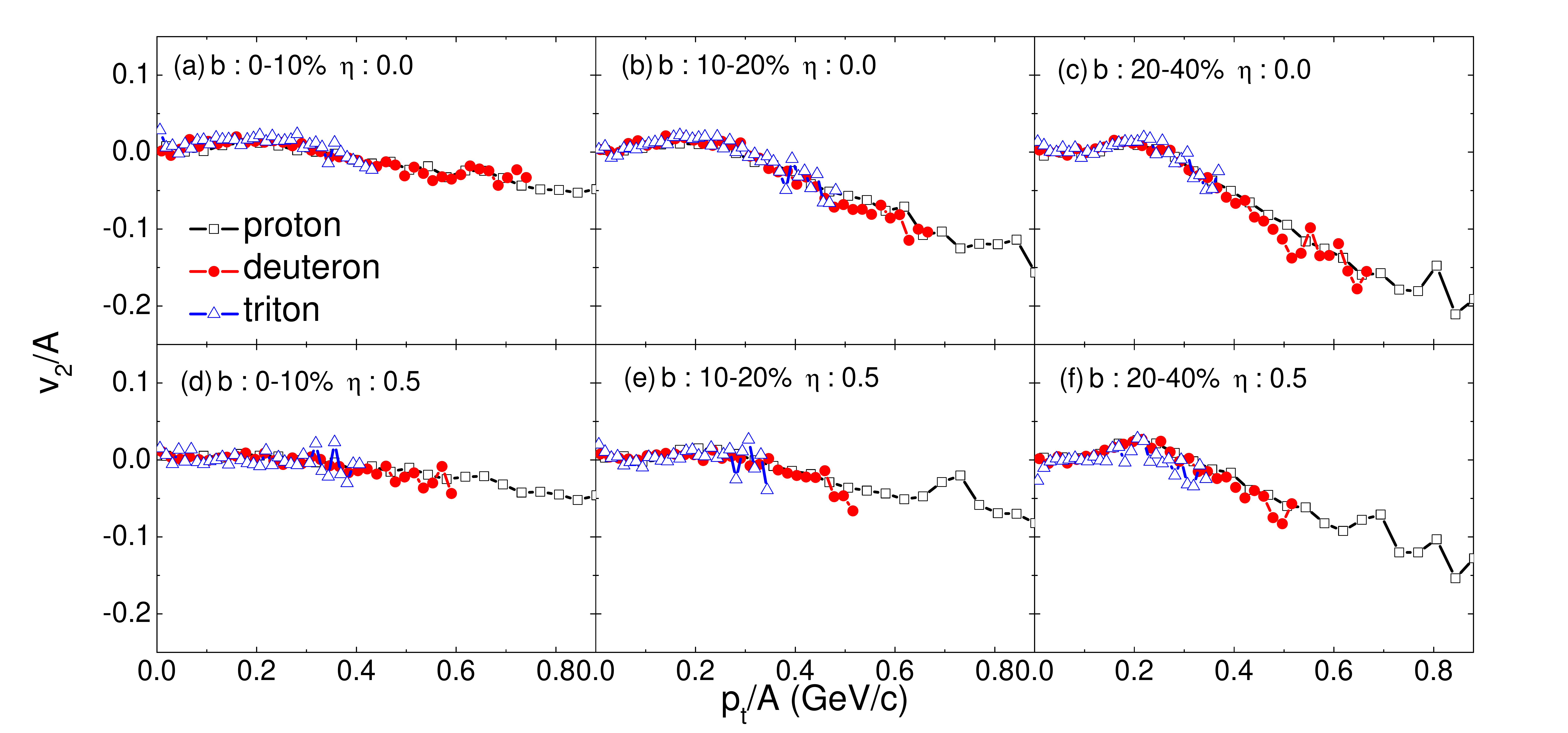}
\caption{(Color online) Same as Fig.~3 but for ${v_{2}}/A$ versus $p_t/A$. 
}
\label{fig4}
\end{figure*}

The panels in Fig.~\ref{fig3} show transverse momentum dependence of elliptic flow $v_{2}$
for light clusters at different centralities with two sets of in-medium nucleon-nucleon cross section. From the figure, it shows that elliptic flows are negative and decrease with the transverse momentum ${p_{t}}$. The negative value of $v_{2}$ reflects that the light clusters are preferentially emitted out of the reaction plane, i.e. so-called squeeze-out phenomenon, and particles with higher transverse momentum tend to be stronger negative  $v_{2}$ values, which is opposite to the collision at lower beam energy~\cite{b21}. In addition, the heavier the clusters, the larger the value of  $v_{2}$. This behavior is similar to the directed flow. In intermediate energy, it is well known that both the attractive mean field and the repulsive nucleon-nucleon collisions play important roles. At lower energy, the mean field becomes dominant in contributing to the formation of a rotating compound system, so the positive elliptic $v_{2}$ is essentially induced by the collective rotational motion~\cite{b16,b41,b42}. In that case, the elliptic flow stems from the attractive mean field. However, at intermediate energy of 0.4$A$ GeV as studied here, the two-body collision gradually plays important role and the collective expansion is therefore developed. 
Although the transverse momentum dependence of $v_{2}$ (the absolute values  of $v_{2}$)  is similar to the previous results at RHIC energies, the mechanism of the flow is very different in the two different energy regions. At RHIC energies, the elliptic flow is mainly driven by the stronger outward pressure ~\cite{b2}, and here the elliptic flow is mainly due to the shadowing effects  of overlapping participant zone. As for the NNCS effect, the larger $\eta$ seems to reduce the absolute  $v_2$ values as in the $v_1$ case. 
After we perform the similar scaling of nucleon number as  $v_{1}$, we observe a nice  number of nucleon scaling exists for the elliptic flow per nucleon as a function of transverse momentum per nucleon as displayed in Fig.~\ref{fig4}. 
In comparison to the number-of-constituent quarks scaling of ${p_{t}/n}$ at RHIC energies, this behavior looks apparently similar in Fig.~\ref{fig4}. The results indicate that the production of light nuclei stems from the nucleonic coalescence mechanism.

\begin{figure*}[!htb]
\includegraphics[width=\hsize]{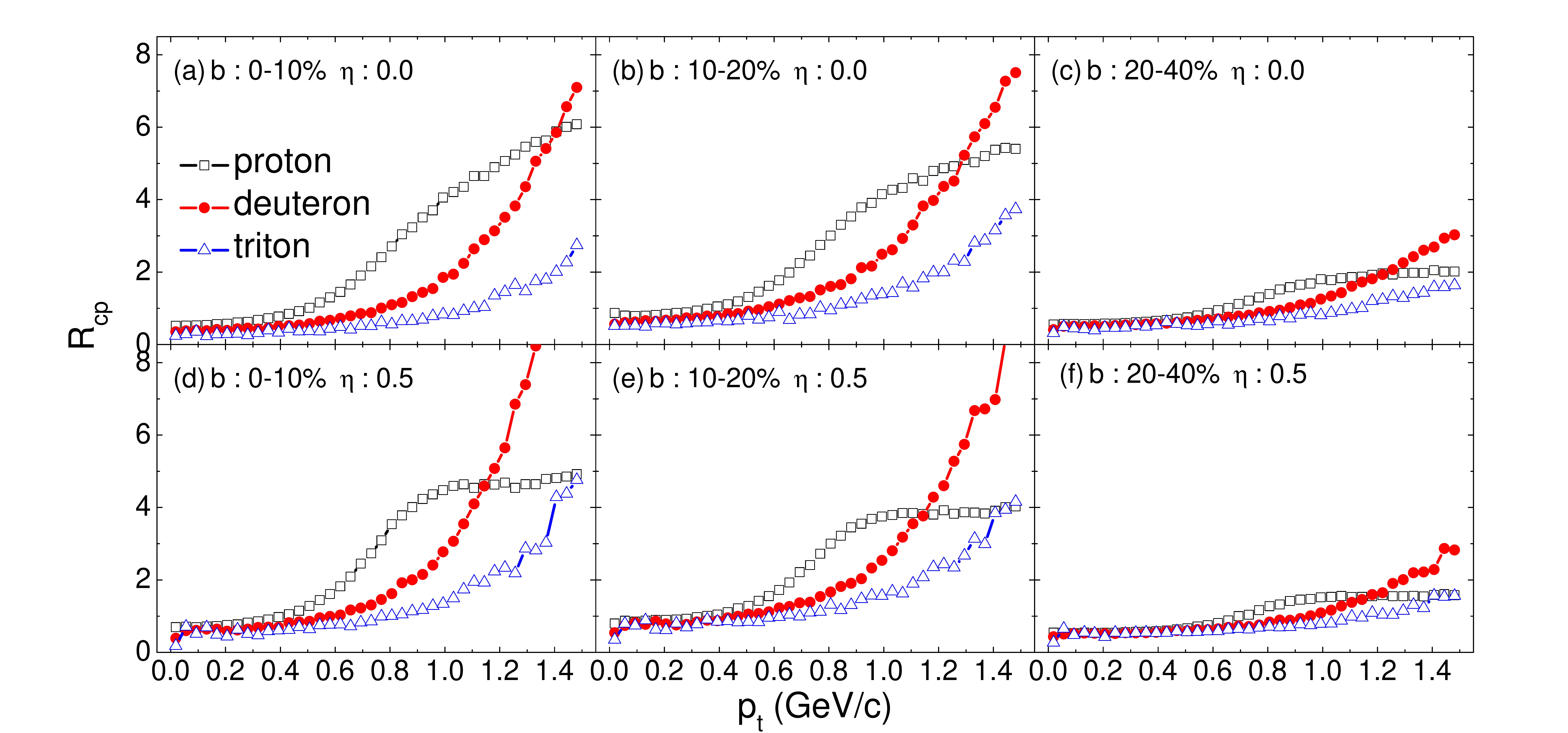}
\caption{(Color online) 
 ${R_{cp}}$ for proton, deuteron and triton in Au + Au collisions at E = 0.4$A$ GeV. From left to right, it corresponds to the ratio of different centralities of 0 - 10$\%$,  10 - 20$\%$,  20 - 40$\%$ to the centrality 40 - 60$\%$. Two cases of the in-medium NNCS are considered: $\eta$ = 0.0 (upper row) and 0.5 (lower row).}
\label{fig5}
\end{figure*}

In Fig.~\ref{fig5} , we show the ${R_{cp}}$ of light nuclei (p,d,t) as a function of ${p_{T}}$ for different central-peripheral pairs in Au+Au collisions at E = 0.4$A$ GeV. It is cleanly seen that the ${R_{cp}}$ curves for proton, deuteron, and triton are obvious different. 
The multiple nucleon-nucleon scattering effect~\cite{b44} tends to transform the longitudinal momentum into the transverse momentum, and the effect becomes stronger with the increase in ${p_{T}}$ in HICs, which leads to larger ${R_{cp}}$ in the high ${p_{T}}$ region. On the other hand, in the low ${p_{T}}$ region, radial flow plays a major role in central collisions, which pushes protons to higher ${R_{cp}}$ regions and results in the smaller ${R_{cp}}$ at low ${p_{T}}$.  For protons, the strength of ${R_{cp}}$ enhancement is suppressed at high $p_T$ with the increase in impact parameter as well as in-medium NNCS factor. We noticed that similar results of ${R_{cp}}$ were obtained in Refs.~\cite{b45,b33} due to the multiple nucleon-nucleon scattering effect or/and the radial flow effect. 

The upper panels and lower panels in Fig.~\ref{fig5} shows the comparison of ${R_{cp}}$ of light nuclei with different $\eta$ values. The $\eta$ values at 0.0 and  0.5 were used in the simulation of Au + Au at 0.4$A$ GeV collisions. In a previous work, the value $\eta$ = 0.2, i.e., 80$\%$ of the free space nucleon-nucleon cross section was obtained~\cite{b46}. In fact, the medium effect is different in various ranges of incident energy and matter density~\cite{b47}. The ${R_{cp}}$ of light nuclei has an increasing trend with ${p_{T}}$ which is explained by the multiple nucleon-nucleon scattering effect as well as radial flow~\cite{b45}, and its trend becomes more rapid for protons with ${p_{T}}$ in the low $\eta$ value case because of the high collision rate between nucleons. Collisions become certainly less in higher $\eta$ values and it makes the multiple nucleon-nucleon scattering  effect less significant. 
Based upon the above argument, the ${R_{cp}}$ of protons is a good quantity for studying the effect of the in-medium NNCS and it means that we can investigate ${R_{cp}}$ to draw a conclusion on the in-medium effect by a quantitative comparison between the model prediction and the data. 

\begin{figure*}[!htb]
\includegraphics[width=\hsize]{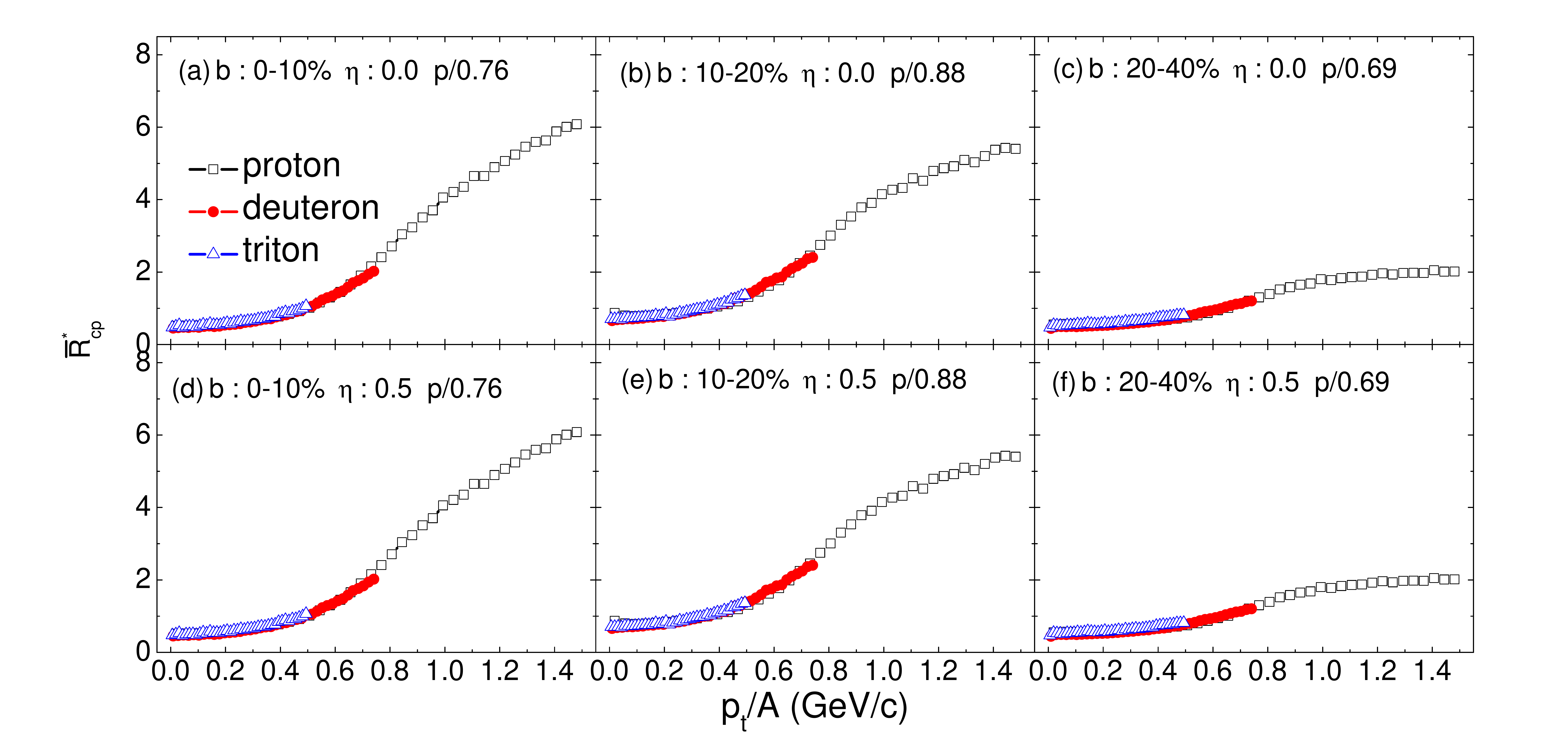}
\caption{(Color online) Same as Fig. 5 but for the scaled ${R_{cp}}$ , i.e. $\widetilde{R_{cp}^*}$.
}
\label{fig6}
\end{figure*}

After performance of the number of nucleon scaling for ${R_{cp}}$, Fig.~\ref{fig6} shows the NN scaling of ${R_{cp}}$ for light nuclei, i.e., $\widetilde{R}_{cp}^{*}$. From this figure, through the number of nucleon scaling, it was found the values of ${R_{cp}}$ for deuteron and triton can be scaled to proton's after using a constant
factor (eg. for panels (a) and (d), a constant factor 1/0.76) on proton's  ${R_{cp}}$.   These  constant factors show no dependence on in-medium NNCS factor and weakly depend on centrality.  All the above phenomena can be seen as the coalescence mechanism for formation of light nuclei. 
 In comparison to the scaling behaviors of ${R_{cp}}$ for light nuclei produced from Pb + Pb at 17.2 GeV, 
Au + Au at 200 GeV, and Pb + Pb at 2.76 TeV ~\cite{b31}, this behavior looks apparently similar in Fig.~\ref{fig6}. The number-of-nucleon scaling  of ${R_{cp}}$ is increasing as a function of $p_{T}$, which could stem from the larger radial flow or multiple nucleon scattering for light nuclei as demonstrated in our earlier work ~\cite{b45}.
Therefore it implies that the number of  nucleon scaling of ${R_{cp}}$ for light nuclei supports the viewpoint of the coalescence mechanism for formation of light nuclei at the kinetic freeze-out stage, 
and this scaling phenomenon is the same as that for elliptic flow ~\cite{b20,b21,b22}.
For the NNCS effect on $\widetilde{R}_{cp}^{*}$, it shows insensitivity due to the cancel effect of the ${R_{cp}}$  ratio.

\section{Summary} \label{sec.IV}

In summary, we have investigated the number of nucleons scaling of anisotropic flows, namely ${v_{1}}$ and ${v_{2}}$ for light charged particles produced by ${^{197}Au} + {^{197}Au}$ at 0.4$A$ GeV where the light particle emission has squeeze-out behavior in the framework of isospin-dependent quantum molecular dynamics model.  The different centralities  and in-medium NNCS factors are taken into account. 
The results show that the curves of both ${v_{1}/A}$ and ${v_{2}/A}$ for different light charged particles stay together, which means that there exists directed flow and elliptic flow scaling on the nucleon number for light charged particles. Then, the number of nucleons scaling of the nuclear modification factor ${R_{cp}}$ for light charged particles also have been  investigated. The results demonstrate that all scaled ${R_{cp}^{*}}$ show a nice overlap with each other after considering a constant difference factor between light charged particles and proton's, indicating a NN scaling of ${R_{cp}}$ is also satisfied.  In light of present study, we propose experimental studies along this direction.

This work was supported by  the National Natural Science Foundation of China under Contract Nos. 11890710, 11890714 and 11421505,  the Key Research Program of Frontier Sciences of the CAS under Grant No. QYZDJ-SSW-SLH002 and the Strategic Priority Research Program of the CAS under Grant No. XDPB09 and XDB16.

\end{CJK*}
\end{document}